\documentclass{article}
\usepackage[font={small,it}]{caption}
\usepackage{algorithm}
\usepackage{PRIMEarxiv}
\usepackage{algpseudocode}
\usepackage[utf8]{inputenc} % allow utf-8 input
\usepackage[T1]{fontenc}    % use 8-bit T1 fonts
\usepackage{url}            % simple URL typesetting
\usepackage{booktabs}       % professional-quality tables
\usepackage{amsfonts}       % blackboard math symbols
\usepackage{amsthm}         % for theorems, definitions, etc.
\usepackage{nicefrac}       % compact symbols for 1/2, etc.
\usepackage{microtype}      % microtypography
\usepackage{amsmath}
\usepackage{lipsum}
\usepackage{fancyhdr}       % header
\usepackage{graphicx}       % graphics
\usepackage{caption}     % multiple figures
\usepackage{subfig}     % multiple figures
\graphicspath{{media/}}     % organize your images and other figures under media/ folder
\usepackage{natbib}
\usepackage{tabularx}

\usepackage[colorlinks = true,
            linkcolor = blue,
            urlcolor  = blue,
            citecolor = blue,
            anchorcolor = blue]{hyperref}

\begin{document}
\onecolumn
%\firstpage{1}
\title{DeepProphet2 - A Deep Learning Gene Recommendation Engine}

\author{
  Daniele Brambilla \\
  TheProphetAI \\
  \texttt{brambilla@theprophetai.com} \\
  %% examples of more authors
   \And
  Davide Maria Giacomini \\
  TheProphetAI \\
  \texttt{giacomini@theprophetai.com} \\
  \And
  Luca Muscarnera \\
  \texttt{lucamuscarnera@gmail.com}
  \And
  Andrea Mazzoleni \\
  TheProphetAI \\
  \texttt{mazzoleni@theprophetai.com} \\
  %% \AND
  %% Coauthor \\
  %% Affiliation \\
  %% Address \\
  %% \texttt{email} \\
  %% \And
  %% Coauthor \\
  %% Affiliation \\
  %% Address \\
  %% \texttt{email} \\
  %% \And
  %% Coauthor \\
  %% Affiliation \\
  %% Address \\
  %% \texttt{email} \\
}
\maketitle

\theoremstyle{definition}
\newtheorem{definition}{Definition}[section]

\begin{abstract}

\noindent Recent advances in the field of machine learning have yielded new tools that show promise for addressing complex problems in the life sciences. The purpose of the paper is to discuss the potential advantages of gene recommendation performed by artificial intelligence (AI). Indeed, gene recommendation engines try to solve this problem: 
\begin{quote}
    If the user is interested in a set of genes, which other genes are likely to be related to the starting set and should be investigated?
\end{quote}
A custom deep learning recommendation engine, DeepProphet2 (DP2), was developed to successfully complete this task. DP2 is available for use by researchers globally, and can be accessed through the website: \href{https://www.generecommender.com?utm_source=DeepProphet2_paper&utm_medium=pdf}{www.generecommender.com}.  Hereafter, insights behind the algorithm and its practical applications are illustrated.

The gene recommendation problem can be addressed by mapping the genes to a metric space where a distance can be defined to represent the real semantic distance between them. To achieve this objective a transformer-based model has been trained on a well-curated freely available paper corpus, PubMed.
The paper provides a comprehensive description of the neural network architecture utilized in the study, including the training process. Multiple optimization procedures were implemented to achieve the optimal balance between bias and variance. The focus was on the impact of factors such as embedding size and network depth. 
In the investigation, the performance of the model in identifying sets of genes related to diseases and pathways was evaluated using cross-validation. The evaluation was based on the assumption that the network had no prior knowledge of pathways or diseases, and that it learned gene similarities and interactions solely through the training process.
Furthermore, to gain a deeper understanding of the gene representation learned by the neural network, the dimensionality of the embeddings was reduced and the results were mapped onto a lower-dimensional space that is easily interpretable by humans. In conclusion, a set of use cases illustrates the algorithm's potential applications in a real-world setting.

\end{abstract}

\section{Introduction}

The task of target identification in life science research is a complex problem: high technical skills and solid expertise are essential to achieve insightful results.

The scope of this paper is to introduce a new approach to this kind of task leveraging the power of data processing that AI makes possible. In this paper, it will be introduced a practical tool developed with this scope in mind that use scientific literature as a knowledge base to suggest to the user other interesting targets for his research. The system, developed for genetic researchers, receives as input a geneset of interest and provides as output other genes, ranked by a similarity score, that relates to the inputs and could be interesting targets.

From a naive perspective, the results of a gene recommender can be seen as the product of the encoding of context-based knowledge in a way that a machine can use to understand the domain of interest \citep{bib:survey_of_embedding_in_NLP}, performing automated reasoning in order to reach meaningful conclusions.

\begin{definition}[Knowledge]
    Knowledge is the ability to link disjoint bits of information and labels \citep{bib:dkiw_paradigm_cognitive_engineering}.
\end{definition}

\noindent
Considering this particular point of view, some initial questions about the problem are paramount to be highlighted:

\begin{enumerate}
    \item[1.] How to effectively encode knowledge in a model that can perform target identification?
    \item[2.] How to understand which kind of reasoning the system performed in order to achieve the results? Are these results coherent with the context?	
\end{enumerate}

\noindent The first question had been addressed by producing multi-dimensional embeddings of the target entity to encode, genes, using the relations among them expressed in the scientific literature. In this way, the encoding is similar to the embedding of words produced by other NLP models like GPT.
The second question has a much more complicated answer, a major deficiency of this type of deep learning models it's their limited explainability, but different tests to investigate the reliability of the results had been made and will be introduced in the validation paragraph. (\ref{sec:validation}).

Within this context, to solve the gene recommendation task, a custom deep learning algorithm has been designed and trained. The structure and the peculiarities of this model, namely, DeepProphet2 (DP2) will be detailed in this document with all the relevant aspects related to its validation. With the aim of making the system available to the research community, the machine learning model has been deployed through a web platform called GeneRecommender, available at \href{https://www.generecommender.com?utm_source=DeepProphet2_paper&utm_medium=pdf}{www.generecommender.com}, freely accessible by researchers worldwide.

The following section of the paper introduces the central concept that forms the foundation of the DeepProphet2 (DP2) model.
A gene recommendation algorithm must understand the subtle underlying relationships between statistical units, trying to mimic the mindset of a researcher, using the computational power of the machine as a tool to make the most unbiased predictions possible. This idea of mimicking a human task is paramount in defining the paradigm that DP2 implements, but it also raises an important new question:
\begin{enumerate}
    \item [3.] How can this human-like reasoning be formalized in a language that a machine can understand?
\end{enumerate}
Following the researcher analogy, the necessary condition for knowledge-based inference is the existence of some kind of structure able to encode this knowledge in an appropriate form. Indeed, the goal is to use this \emph{a priori} information to draw conclusions.

Pursuing this path, a variation of the Transformer architecture \citep{bib:attention_is_all_you_need} was developed.
Transformers are a new class of deep neural network models that have emerged in recent years in the context of Natural Language Processing (NLP). They are based on the substructures of encoders and decoders, which are - from the point of view of information theory - the implementation of functions that contract and expand the information in order to extract meaning from the input. These functions have to be learned from data \citep{bib:understanding_autoencoders}.
This type of information manipulation occurs in conjunction with the mechanism of attention, which aims to create a machine learning equivalent of the human ability to focus on concepts that are relevant in a particular context \citep{bib:understanding_human_and_machine_attention}.  The ability to extract meaningful information and, heuristically speaking, contextualize it (both in the single ``phrase'' intended as the combination of genes in the query and in the ``vocabulary'' intended as the whole pool of genes) produces powerful models, able to perform complex reasoning on input data.

However, from a technical perspective, the model here presented is not a Transformer in the most classical sense. 
Indeed, the decoder part of this model is reduced to a linear algebra manipulation of the encoder output, thus resembling more the \emph{Generative Pre-trained Transformer} (GPT) model category \citep{bib:gpt} rather than the one described in the original paper.
This methodology enables to produce an information contraction of the input data which is clearly interpretable and explanatory of the obtained output since the relationship between encoding and output is given (and not a function that has to be learned from data).
Moreover, this approach reduces the overall number of free parameters of the network, improving its ability to generalize and reducing the variance induced by over-parametrization \citep{bib:effects_of_overparametrization}.

This particular approach proved extremely useful in the hyper-parameter tuning phase: a sub-optimal value for the dimension of the embedding space was determined in order to achieve a meaningful contraction of the information explaining the variance of the different phenomena.

\section{Material and Methods}
\subsection{Network Topology}

\noindent In order to approach the task outlined above, a large part of the research activity has been devoted to the search for an optimal topology of the neural network.
The role of an optimal topology can be decomposed into two sub-tasks: 

\begin{enumerate}
    \item[1.] Representation of genes and disease groups
    \item[2.] Analysis of different representations of genes.
\end{enumerate}

\begin{figure}
  \centering
  \includegraphics[width=1.0\textwidth]{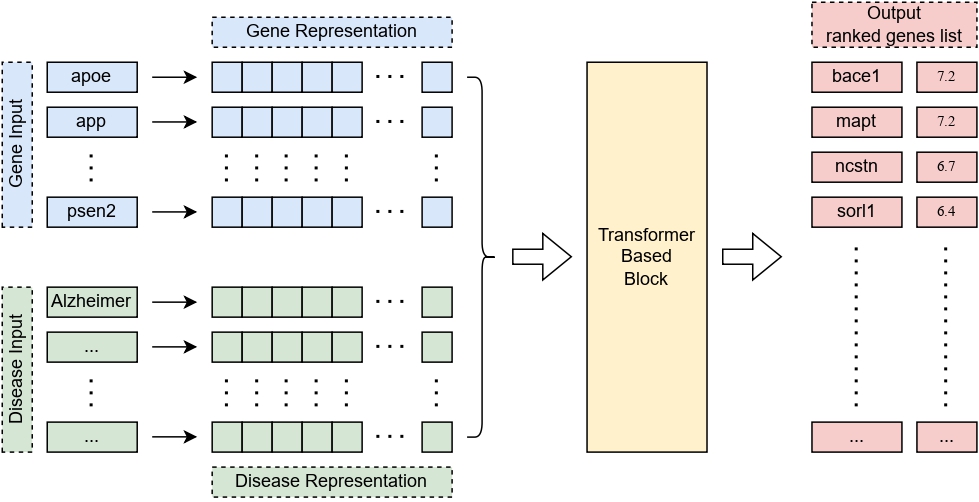}
  \caption{Visual representation of the topology of the network.}
  \label{fig:topology}
\end{figure}

\noindent The former is a non-trivial problem. The representation must capture information about the role of each element in the set and, using a data-driven approach, find the right way to extract the meaning of each gene or disease.
To achieve this result, the mechanism of self-attention has been used. As described in transformer's original paper \citep{bib:attention_is_all_you_need}, self-attention provides a tool for the network to understand the links between different concepts (such as genes and diseases) that is fully learned as network parameters, therefore representing a very versatile approach.
In terms of analyzing the different representations, again the transformer has been used as the building block for the network. Using the transformer block, the network (Fig.\,\ref{fig:topology}) is able to produce a ranked list of genes ordered by semantic distance from the input query.
This ranking list is created using an attention mechanism based on the scalar product as a measure of similarity. This idea, analogue to the concept of cosine distance, provides a computationally efficient way to measure similarities between entities. 
It is interesting to note that this ranking task fully pursues the goal of the network. In the following chapters, the correspondence between the ability to rank genes and the study of their interactions will be analyzed in detail.

\subsection{Network Structure}
\label{sec:network_structure}

\noindent The network structure proposed in this work consists of three main parts: embedding, encoder, and decoder. 

The embedding part is responsible for creating an abstract representation of genes and diseases, which is achieved through embedding layers that are randomly initialized. The embeddings are learned during the training phase and are not generated by external algorithms or pre-trained.

The encoder part of the network learns correlations and similarities between the input gene and disease embeddings exploiting the attention mechanism, analogously to the transformer architecture described in the GPT paper from OpenAI \citep{bib:gpt}. 

The decoder on the other hand is analogous to a softmax classifier among all genes. It operates by computing a matrix product between the last transformer layer output and the embeddings lookup table, with a softmax activation function applied at the end to obtain a probability distribution. No negative sampling techniques are employed, but the whole gene embedding matrix is used in the product.

The network training is performed in an unsupervised manner, analogously to that described in the \citep{bib:gpt} paper. Given an article with $N_g$ gene citations and $N_d$ disease citations, at each step, the first $N_g - 1$ genes and all $N_d$ diseases are provided as input, with the last $N_g - 1$ genes being provided as the target. A mask is applied to the transformer attention layers, allowing information to only flow from the end of the input list to the start of it, thus masking the target from the input and performing a virtual data augmentation process at the same time. The categorical cross-entropy loss is then computed and used for optimization with the Adam algorithm. 

This training procedure is again inspired by the one described in \citep{bib:gpt}. For a more formal description, the reader is referred to the paper itself, while a more intuitive explanation can be found in section \ref{par:training_and_data_aug}.

In terms of model size, the embedding dimension is set to $64$ for both genes and diseases, with embedding lookup tables of size approximately $25\,000$ for genes and $11\,000$ for diseases, depending on the exact number of distinct genes and diseases in the training set. The transformer block is made up of $6$ layers with $8$ attention heads, totalling around $300\,000$ free parameters. The total number of parameters in the network is therefore approximately $2.5$ million.

\subsection{Mathematical and Heuristic Assumptions on the Gene Recommendation Task}

\label{section:mathematical}
\noindent This implementation of the transformer architecture is based on a simple assumption: since the scalar product produces a measure of similarity in a given space equipped with the inner product, it is necessary to develop a function that manipulates the data projecting them in a certain space, such that the scalar product is also a measure of similarity among genes.
Formally, the idea is to produce a network able to learn a function – more specifically an isomorphism \citep{bib:conceptual_mathematics_cap2_isomorfismi} – able to map genes in a Hilbert space where the scalar product is consistent with the semantic distance among the genes \citep{bib:efficient_representation_low_dim_manifolds_using_dn}\citep{bib:morphism_is_all_you_need}.

\begin{definition}[Hilbert Space]
    A Hilbert space is a vector space H equipped with a scalar product such that H is complete for the norm~\citep{bib:brezis}.
\end{definition}

\noindent The notion of semantic distance, here introduced, finds its foundations in the assumption that it is possible to quantify the strength of the relationship between genes in terms of interaction in biological processes of the proteins they encode.

\begin{definition}[Semantic Distance]
The\emph{semantic distance} is defined as a function $\sigma:\ G\ \times\ G\ \rightarrow\mathbb{R}^+$ such that the following conditions are satisfied:\\
\begin{align}
\sigma( g_1, g_2 ) & < \sigma(g_1,g_3) \iff g_1 \text{ relates with } g_2 \text{ more than } g_3 \,\forall \{g_1, g_2, g_3\} \in G \\
\sigma( g_1, g_2 ) & = \sigma( g_2, g_1 ) \\
\sigma( g_1, g_1 ) & = 0  \\
\sigma( g_1, g_2 ) & + \sigma(g_2, g_3 ) \ge \sigma(g_1,g_3)
\end{align}

\end{definition}

\begin{figure}
  \centering
  \includegraphics[width=\textwidth]{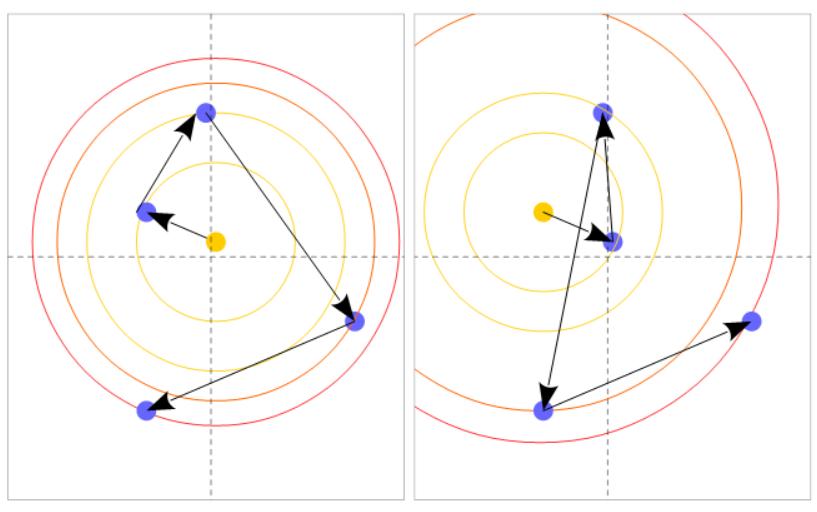}
  \caption{Visualization of order relationship induced by a distance.
  In this example, the order relationship is defined according to the euclidean distance from the point that in the figure is highlighted in orange. This is possible because distances are defined in the right semi-axis of the real numbers line, which is equipped with an order relationship.
  }
  \label{fig:order_relationship}
\end{figure}

\noindent Assuming that such a function could exist and be well defined, it would induce an order relationship for every gene in the dataset (Fig.\,\ref{fig:order_relationship}). 
Given a gene $g_j \in G$, it is indeed possible to define the following order relationship:
\begin{equation}
    g_i \le_j g_k \iff \sigma(g_i,g_j) \le \sigma(g_k,g_j) \forall g_i, g_k \in G .
\label{eq_01}    
\end{equation}

\noindent The optimization problem that describes the training of the network can then be then formulated.

\begin{definition}[Optimization Task]  
Let $X$ be an Hilbert Space. Find $\phi : G \rightarrow X$ such that: \\
\begin{equation}
        \langle {\phi(g_i),\phi(g_j)} \rangle \ge \langle {\phi(g_k),\phi(g_j)} \rangle \implies g_i \le_j g_k 
        \quad
        \forall {g_i,g_j,g_k} \subseteq G
\label{eq_02} 
\end{equation}
being $\langle\cdot\,,\cdot\rangle$ the Hilbert space scalar product.
\end{definition}

\noindent This method offers a different perspective on the task by generating a clear geometric representation of the relationship among genes: namely, it introduces the idea of orthogonality between genes, which results in genes with a larger semantic distance having a smaller inner product in the projection space \citep{bib:geometry_of_culture} \citep{bib:theoretical_foundations_and_limits}.

The objective of this optimization problem is to produce a suitable representation of genes in the Hilbert
space.
Indeed, assuming that a certain representation of the genes that does not fulfil the optimization problem, for example, a random distribution of vectors in the Hilbert space such that every vector is associated with one and only one gene.
It is straightforward to observe that, if the vectors are chosen randomly, there is no way to recover the association with genes by only observing the set of vectors in the Hilbert space, since there is no
information to recover.
Starting from this statement, there is an opposite point of view: an objective is to exploit the properties of the Hilbert space in order to store as much information as possible about the genes.
In fact, if it is possible to preserve information about genes, it is also possible to study their representation in order to understand them.
In particular, the inner product of the Hilbert space offers a powerful tool to represent this information.
The optimization task aims to this objective: finding the best arrangement of points in this space such that the inner product preserves as much information as possible about genes, which in this framework consists of the set of semantic distances between them.
In practical terms, preserving the information translates into having the possibility of performing analysis on the projected data.
In fact, if data in Hilbert space represents in an efficient way real life data, then it is possible to use it for inference.
The proposed model uses this idea exploiting the inner product in the attention mechanism, which relies heavily on the possibility of measuring similarity through the inner product.

\begin{figure}
  \centering
  \includegraphics[width=0.7\textwidth]{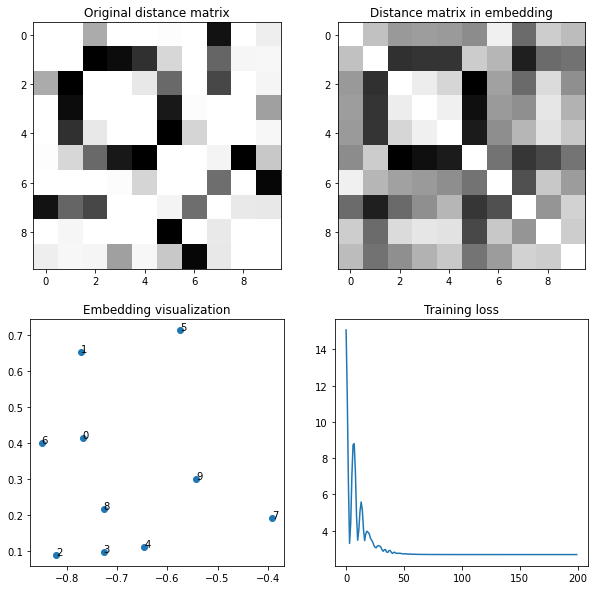}
  \caption{A procedure similar to the one described in section\,\ref{section:mathematical} is applied on a random distance matrix. The product of the algorithm is the embedding (left-bottom) and the reconstructed distance matrix (right-top) which show the preservation of the inner structure of data in the projection space. It is possible to imagine all the concepts described in this chapter as strictly related to this simple visualization, with the further task of learning what the original distance matrix is.}
  \label{fig:embedding_algorithm}
\end{figure}

Heuristically, the role of this geometrical approach is based on the idea of producing results that encode information in a certain shape. Indeed, the geometry of the map of genes in this new Hilbert space (that from now on will be denoted with the term \emph{embedding}) allows the user of the system not only to perform inference in the standard sense, but also to study the geometry and the topology of projected data. In fact, the system allows the researcher to have a visual representation of how genes behave opening to different kinds of analysis exploiting the structure of the embedding which, thanks to its low dimensionality, could be used as a refined source of data to perform statistical learning on genes (e.g. unsupervised learning algorithms such as clustering or supervised algorithms such as KNN, LDA\dots) \citep{bib:empirical_comparison_between_autoencoders}.
For an easier understanding, Fig.\,\ref{fig:embedding_algorithm} shows an application of the procedure described in this chapter on a simpler toy model. 

\subsection{Training and Data Augmentation}

\label{par:training_and_data_aug}

\noindent In order to produce a robust model the team tried to capture the dynamics of gene behaviour by creating statistical units composed of relations between diseases and genes. This process was performed using an NLP application optimized for medical research – the online service PubTator \citep{bib:pubtator} \citep{bib:taggerone} – to analyze portions of medical research papers and tagging entities like genes and diseases. In this process, all genes which are not present in the human genome are dropped, so DP2 has only the notion of human genes (the possibility to consider also genes of other species is under development). Every statistical unit of the training set then consists of an article composed of a set of genes and diseases, and it is the result of the application of PubTator on each PubMed article. Co-citation within documents is used as a proxy for determining the presence of a relationship between two genes. No further characteristic of the relationship has been considered (i.e. up or down-regulation). The NLP engine used had proved good reliability with an F1 score  of $0.867$ for the gene tagging task and $0.837$ for the disease tagging task.

The dataset resulting from the application of the described NLP engine can be directly downloaded from the \href{https://ftp.ncbi.nlm.nih.gov/pub/lu/PubTatorCentral/}{PubTator} website. Among all biological entities tagged, only genes and diseases were used in training, resulting in a pool of $7\,226\,607$ articles. The number of distinct genes amounts to $25\,052$ and the distinct diseases to $11\,655$, for a total of $34\,560\,857$ gene citations and $67\,490\,174$ disease citations.
Genes are codified by the NCBI standard \citep{bib:NCBI} and MESH terms \citep{bib:MESH} are used for diseases.

Due to the complex nature of the phenomena under study, data has been manipulated with a particular procedure of data augmentation \citep{bib:data_augmentation} in order to maximize the information that the network can use to learn.
Indeed, the training has been enhanced by producing a virtual expansion of the dataset, embracing the idea of inducing clear relationships between every single possible couple of genes, in a way that recalls the notion of semantic distance defined before.
Following what has been proposed in the first paper relative to the GPT family of models \citep{bib:gpt}, the following technique was applied  \citep{bib:unsupervised_pretraining_for_seq_to_seq}.

\begin{definition}[Data Augmentation Procedure]  
Let $G_a = \{g_1,g_2,...,g_p\} \subseteq G$ be a set of genes with a known relationship. It is possible to define:
\begin{equation}
    G_{a,i} := \{ g_j \in G_a : j \le i \} \ \forall \ i \in \{1,2,...,p\}.
    \label{eq_03} 
\end{equation}
\end{definition}

\noindent This kind of data augmentation allows the network to understand the role of every single gene, refining the granularity of the training data in order to understand genes as single entities and not only as part of sets, also preserving the cluster-like structure induced by the semantic distance.

It is important to make a clear distinction between this kind of data augmentation and other statistical techniques such as oversampling. 
Oversampling (and affine statistical methods such as bootstrap) is a technique that tries to compensate for the unbalance in sample categories by simply refeeding part of the data during the training phase.
In the case that is being presented, it may appear like some kind of oversampling is being performed, but it is important to denote how the sample, in this context, is not the gene \emph{per se} but instead the sequence where the gene is presented.
In other words, taking sub-sequences of the original data allows presenting to the network different observations of the same phenomenon. Training the network to process partial information about the same object helps the network to understand the underlying concepts that explain a certain biological behaviour.

Moreover, a peculiarity of the presented data augmentation is that this algorithm is built \emph{inside} the network itself.
While often the data augmentation phase is an extra step to perform before the training in order to build a richer dataset, here it is incorporated in the network through a masking technique (see section \ref{sec:network_structure} for details) in order to allow a more efficient training.

This whole process was carried out taking inspiration from the philosophy of curriculum learning \citep{bib:curriculum_learning}: the aim was to extract information in the same way as a biomedical researcher so that the system could learn the relationships between genes by managing the flow of information appropriately. 
Following this idea, the following information extraction strategy was developed:

\begin{enumerate}
    \item [1.] Papers are subject to a threshold interval in terms of genes cardinality: papers that contain too much information are also associated with high variance data and could therefore lead to information quality pollution. On the other hand, papers that do not contain enough information are subject to a high risk of bias and are therefore not robust enough to be used.
    \item [2.] Full-Text\,/\,Abstract mining strategy: biomedical publications are considered reliable sources of information thanks to the rigorous quality control they have to pass in order to be published. Therefore, information is extracted from all available text areas of the article, using the full paper when possible or making the NLP engine focus on the abstracts when the full text is not available \citep{bib:generating_summary_for_scientific_paper_review}. 
\end{enumerate}

\noindent As the reader may have noticed, the strategy is intended to optimize the bias-variance tradeoff. Natural Language is a high noise data source~\citep{bib:dealing_with_noise} and these methods are built to produce high quality information from this kind of source \citep{bib:deep_learnin_models_are_not_robust}.  

Along with data generation constraints, the focus-like mechanism induced by self-attention has been exploited: referring again to the previous analogy, the mechanism of attention is built to avoid the ideal researcher to merely memorising data or, more technically, to avoid the over-fitting phenomenon. As for the hyper-parameter tuning phase, the simple structure of the decoder allowed to produce better results by reducing the variance of the model and optimizing the risk induced by the high noise of the data (since optimizing the hyperparameters is useful in reducing the risk of overfitting).

In order to measure the ability of the model to explore and process information from the training data, a validation procedure was implemented.
In the context of embedding-based networks, validation of results plays a key role: if on one hand, it provides a powerful tool to quantify the efficiency of the network in effectively encoding knowledge, on the other hand, it allows to test of the behaviour of the system in the main usage contexts.

% For Original Research articles, please note that the Material and Methods section can be placed in any of the following ways: before Results, before Discussion or after Discussion.

\section{Results}

\subsection{Validation}
\label{sec:validation}

\noindent From a computational learning theory perspective, the notion of bias-variance trade-off in the context of embedding-based networks is opaque, but it can be considered, with a heuristic approach, as the optimal point between the model having a too shallow point of view on the phenomena of study (high bias) and the model simply memorizing the inner structure of the data (high variance).

For example, if the space where the genes are projected is too small the model still captures part of the dynamics of the data, but it cannot give a detailed explanation of those dynamics. On the other hand, if the embedding space is too big, the dimensions of the projection space start to become poor in terms of explained information.

\subsubsection{Validation Procedure}
\label{par:validation-procedure}

The task that the DP2 model addresses can be assimilated to a multi-class classification problem, thus techniques and tools from this field can be used to obtain a model validation. In particular, a leave-one-out cross-validation was performed using a one-vs-rest (OVR, a.k.a. one-vs-all, OVA) approach to transpose the problem from multi-class to binary classification. The metric of choice was the Receiver Operating Characteristic curve (ROC) and the relative area-under-the-curve (AUC), to disentangle results from the choice of the point-of-work. In the following lines, the reader can find a detailed description of all the phases of the validation process, along with a reference to the results.

First of all, a gene set is identified as an ensemble of genes with a common factor, e.g. taking part in the same biological process or pathway, or being related to the same disease. Then, leave-one-out cross-validation is performed, as described by algorithm \ref{alg:crossvalid}: the gene set is split into two parts, a target one composed of only one gene, and an input one including all the remaining genes. Then, model predictions are computed for the input set, and a Confusion Matrix (CM) is obtained with an OVR approach: a binary CM is computed for every single output class (in this case, one for every output gene) and an overall sum CM is computed to get a measure of the average model performance among genes. This process is then repeated for every possible input-target combination and the obtained CMs are summed together to get a measure of the model performance on the selected gene set. All this is done by applying a variable threshold to the model output.
Using the obtained gene set specific CM, true-positive-rate and false-positive-rate are computed as
\begin{align}
    tpr = \frac{tp}{p} &= \frac{tp}{(tp+fn)} \\
    fpr = \frac{fp}{n} &= \frac{fp}{(fp + tn)}.
\end{align}
Both values are computed for each selected threshold and can be used to build the ROC curve and compute the AUC.
At the end of the process, an average of the metrics can be computed to obtain insights on the model performance on the whole dataset, and thus get an estimate of the knowledge of the model itself in the related field.

\begin{algorithm}
\caption{Cross Validation Procedure}\label{alg:crossvalid}
\begin{algorithmic}
\State $p \gets \{g_1,g_2,...,g_n\}$  \Comment{Gene set initialization}
\State $subset_p[i] \gets p / \{g_i\} $  \Comment{Definition of the subset family}
\State $test_{res} \gets [\empty]$
\For{$i \in [1,n]$}  
    \State $res \gets predict(subset_p[i])$  \Comment{Prediction of the $\text{i}^\text{th}$\,subset}
    \State $test_{res} \gets [test_{res},res ]$  
\EndFor
\State $return \longleftarrow eval( test_{res} )$
\end{algorithmic}
\end{algorithm}

\subsubsection{Benchmark Datasets}

Having defined a general validation procedure based on gene sets, it is interesting to evaluate the performance of DP2 on a set of different contexts. 
Being trained on literature, the model should be able to extract information from a wide variety of topics covering the majority of the possible research applications.

For this purpose, a set of online available databases was identified, with the aim of using the contained data to build sets of genes related by a given concept or biological entity. In the next paragraph, the reader can find a list of the identified sources, together with references and a brief description.

\paragraph{Reactome}
Reactome \citep{bib:reactome} is an open-source, open access, manually curated and peer-reviewed pathway database. 
A pathway is defined as a series of connected reactions between proteins, with each individual reaction considered as a single step in the pathway. Two or more proteins take part in every reaction. 
This data can be used to sets of genes belonging to the same pathway.

\paragraph{Gene Ontology}
The Gene Ontology (GO) knowledge base \citep{bib:geneontology1} \citep{bib:geneontology2} is the world’s largest source of information on the functions of genes. This knowledge is provided in both human-readable and machine-readable form, and is considered a foundation for computational analysis of large-scale molecular biology and genetics experiments in biomedical research.

In particular, three GO classifications exist:
\begin{enumerate}
    \item \emph{Biological Process} The larger processes, accomplished by multiple molecular activities (note that a biological process is not equivalent to a pathway).
    \item \emph{Cellular Component} The locations relative to cellular structures in which a gene product performs a function.
    \item \emph{Molecular Function} Molecular-level activities performed by gene products we use lists of genes related to the same class term as gene sets for the evaluation of our algorithm
\end{enumerate}
Gene sets can be defined starting from genes belonging to the same GO class (i.e. the same biological process, cellular component or molecular function).
    
\paragraph{DISEASES} DISEASES \citep{bib:diseases} is a weekly updated web resource that integrates evidence on disease-gene associations from automatic text mining, manually curated literature, cancer mutation data, and genome-wide association studies. Three different types of gene-disease association datasets are provided:
\begin{enumerate}
    \item \emph{Knowledge} disease-gene associations from curated sources.
    \item \emph{Experiments} disease-gene associations inferred from experiments data.
    \item \emph{Textmining} gene-disease associations extracted from co-citation data produced by automatic text mining of the biomedical literature.
\end{enumerate}
Lists of genes associated to the same disease can be used as gene sets for algorithm validation.

\paragraph{STITCH}
STITCH \citep{bib:stitch} is a database of known and predicted interactions between chemicals and proteins. The interactions include direct (physical) and indirect (functional) associations; they stem from computational prediction, from knowledge transfer between organisms, and from interactions aggregated from other (primary) databases.
Genes interacting with the same chemical can be used to build gene sets for algorithm validation. In particular, STITCH provides a confidence score for each interaction, spanning values from $0$ to $999$. Only interactions with score $s > 900$ will be considered, corresponding to what STITCH defines as ``highest confidence'' matches.  

In general, for each of the selected sources, only gene sets with a number of elements between $5$ and $200$ are included in the validation set.

\subsubsection{Validation Results}
\begin{figure}[htpb]
  \centering
  \includegraphics[width=.7\textwidth]{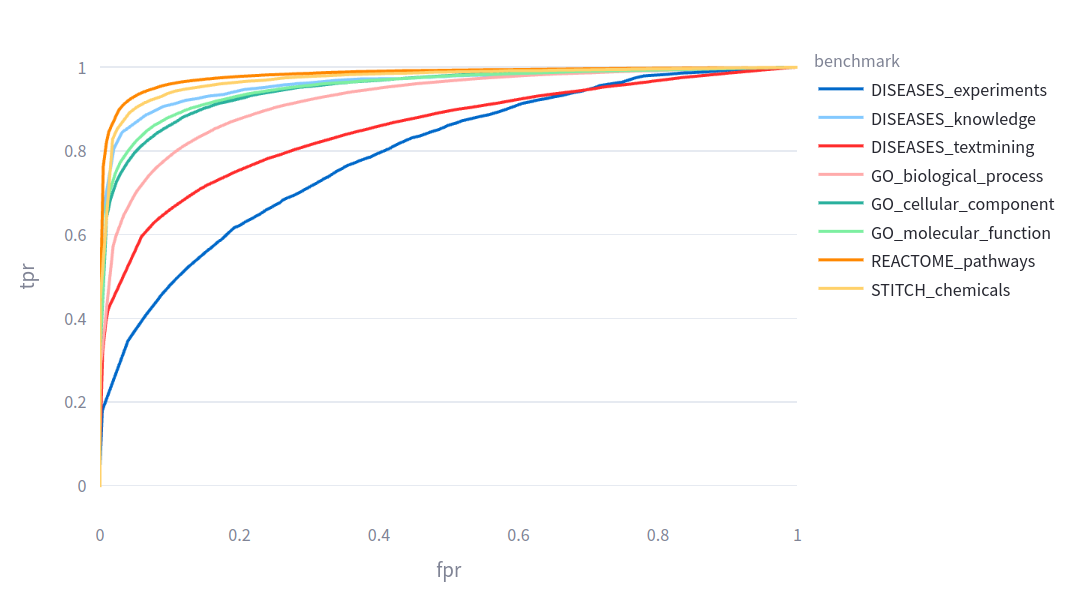}
  \caption{ROC curves obtained with the validation procedure described in Section \ref{par:validation-procedure} for the DeepProphet2 algorithm on all the listed benchmark datasets.}
  \label{fig:roc_multidataset}
\end{figure}

In Fig\,\ref{fig:roc_multidataset} the ROC curves resulting from the evaluation process on all benchmark datasets are reported. Area Under The Curve (AUC) values can be found in Tab\,\ref{tab:validation_results}, first line. As can be seen, the DP2 algorithm performs very well for pathways and chemicals, showing an underlying partial understanding of the interactions between proteins or with other chemical compounds. 
Notably, also the \emph{DISEASES knowledge} benchmark exhibits a very high AUC metric, probably due to the fact that the model is trained with both genes and disease terms, and the gene-diseases matches in this subset being very accurate and well known in the literature.

Good scores are obtained for the Gene Ontology classes as well, showing that the model can extract common properties like molecular function or cellular component from the input genes set, and expand the set with other genes with similar properties.

The worst results are obtained in the DISEASES Experiments and Text mining case, probably due to the intrinsic difficulty of the prediction of disease-gene associations not yet well consolidated in the literature.

\subsection{Performance Framing and Comparison}

No reference to deep learning application to the gene recommendation task could be found in the exsisting literature, making comparison with the state of the art a non trivial task.
Despite this, it is still possible to build alternative algorithms with the same input/output signature of the DeepProphet2 model. In such a way, the DP2 performance can be compared with simpler solutions to the same problem, providing a frame of reference in which the predicting power of the algorithm can be more easily assessed.

In particular, alternative algorithm should present the following common structure:
\begin{itemize}
    \item Input should be represented by a list of genes.
    \item The algorithm should output a score $s\in[0,1]$ for every existing gene.
    \item Genes in the input list should not appear in the output list (their score should be 0).
\end{itemize}

Then, the same validation procedure (see section \ref{par:validation-procedure}) can be applied to produce a comparison between all the considered algorithms.

Follows a description of the three different reference algorithms that have been identified for the scope.

\paragraph{Random}

This is a model that performs random gene recommendations: a random score between $0$ and $1$ is provided for each gene, completely ignoring the provided input genes.
This algorithm is provided as a double check that the validation process is always working as expected. The ROC curve in particular is expected to be a straight line from $(0,0)$ to $(1,1)$, with an AUC of $0.5$.

\paragraph{Frequency Based}

This model provides recommendations based on the number of articles citing each gene in the training set (i.e. in the scientific literature). The score is computed as follows:
\begin{equation}
    \mathcal{S}_j := 
    \begin{cases} 
        0   &\text{if } g_j \in \mathcal{G}_i \\
        \nicefrac{f_j}{\max_{j}(f_j)}  &\text{otherwise} \\
        
    \end{cases}
\end{equation}
where $\mathcal{G}_i$ is the input genes set, $f_j$ equals the number of articles in which a gene $g_j$ is cited in the training set. 
In this way, taking the N top scored genes, the distribution of recommended genes matches the frequency distribution of the training dataset.
This algorithm is provided as a reference baseline: all the other methods are expected to produce better results to be considered useful in a real case scenario.

\paragraph{Nearest Neighbour (NN) Search}

Given a method to map every gene to a vector representation in a generic embedding space, a recommendation algorithm can be constructed by means of similarity between input genes embeddings and all the other genes embeddings.

Given
\begin{align}
    \begin{split}
        \mathcal{G} &:= \{\textbf{g}_1, \dots, \textbf{g}_i, \dots, \textbf{g}_K\} \\
        \mathcal{E} &:= \{\textbf{e}_1, \dots, \textbf{e}_j, \dots, \textbf{e}_N\}
    \end{split}
\end{align}
where $\textbf{g}_i$ and $\textbf{e}_i$ are vectors in the genes embedding space, K is the number of input genes and N the number of existing genes, a score can be defined as
\begin{equation}
    \mathcal{S}_j := \frac{1}{K} \sum_{i=1}^{K} \frac{\mathcal{C}(\textbf{g}_i,\textbf{e}_j)+1}{2}
\end{equation}
where $\mathcal{C}$ is the standard cosine similarity metric, which lays in the interval $[-1,1]$. In this way $S_j \in [0,1]$ is a normalized ranking score for all genes.

Different algorithms can be used to generate gene embeddings. One possible choice is the use of transformer models pre-trained on biomedical text, which are suitable for creating word embeddings for terms in the biomedical domain, including gene symbols. In this study, three different models have been identified based on the current literature and state of the art in the field:

\begin{itemize}
    \item BioBERT \citep{bib:biobert}, a pre-trained language representation model specifically designed for biomedical text mining. It utilizes the BERT architecture and is pre-trained on large-scale biomedical corpora, outperforming previous state-of-the-art models on multiple biomedical text mining tasks such as named entity recognition, relation extraction, and question answering. The pre-trained weights of BioBERT are freely available along with the source code for fine-tuning.
    \item SapBERT \citep{bib:sapbert}, a pre-training scheme for biomedical entity representations, aimed at accurately capturing fine-grained semantic relationships in the biomedical domain. It offers an elegant one-model-for-all solution to medical entity linking by self-aligning the representation space of biomedical entities, achieving a new state-of-the-art on six MEL benchmarking datasets and outperforming various domain-specific pre-trained MLMs such as BioBERT\citep{bib:biobert}, SciBERT\citep{bib:scibert} and PubMedBERT\citep{bib:pubmedbert}. The framework utilizes a scalable metric learning framework that leverages UMLS\citep{bib:umls}, a massive collection of biomedical ontologies with 4M+ concepts.
    \item KRISSBERT \citep{bib:krissbert}, a biomedical entity linker that utilizes Knowledge-Rich Self-Supervision (KRISS) to overcome the challenges faced by standard classification approaches and zero-shot entity linking. It generates self-supervised mention examples on unlabeled text using a domain ontology and trains a contextual encoder using contrastive learning, and maps the test mention to the most similar prototype for linking. KRISSBERT attains new state of the art, outperforming prior self-supervised methods on seven standard datasets spanning biomedical literature and clinical notes, without using any labeled information.
\end{itemize}

All the employed model are available on \href{https://huggingface.co/}{huggingface.co} through the Feature Extraction hosted inference API.

\subsubsection{Results}
\begin{figure}
  \centering
  \includegraphics[width=.7\textwidth]{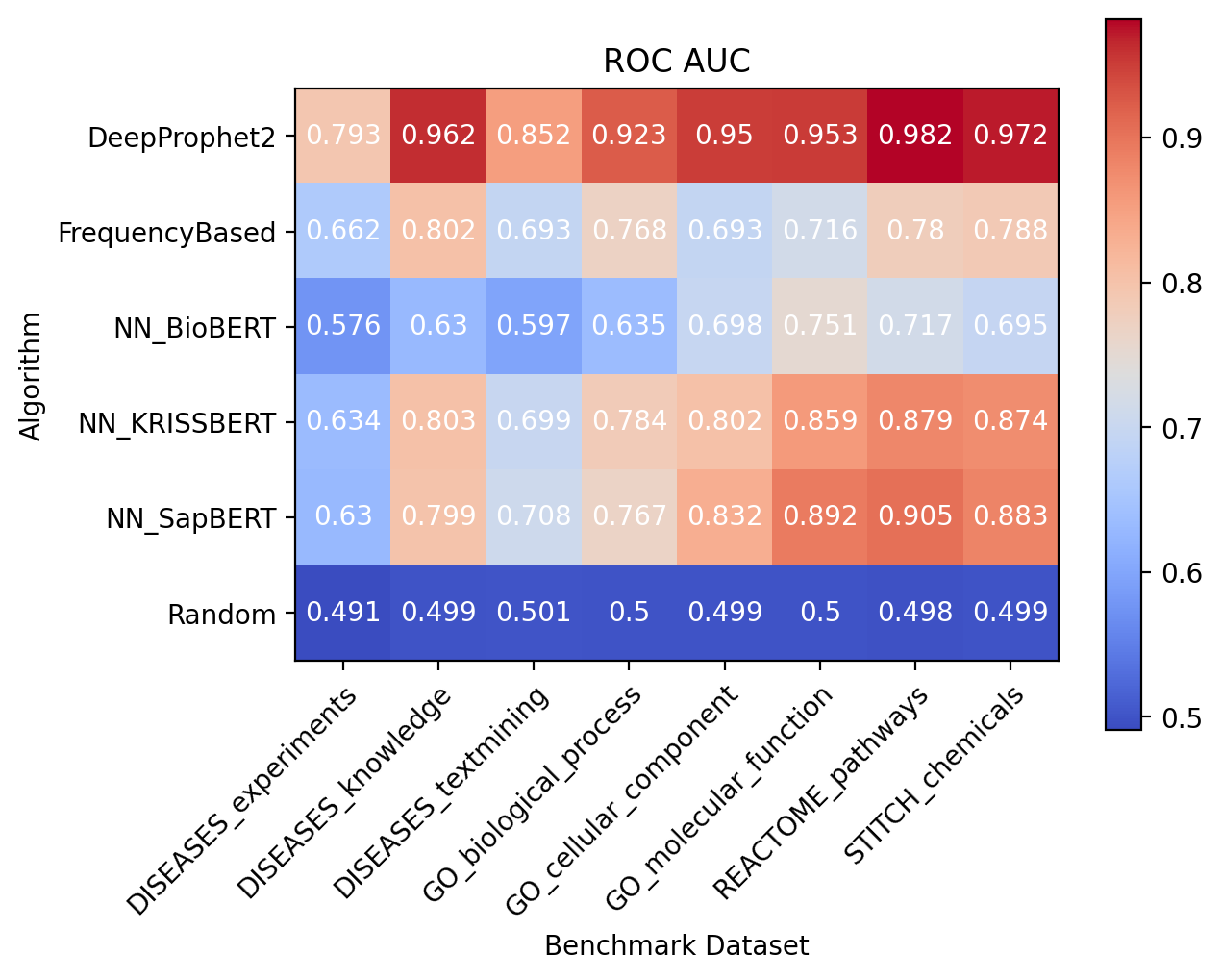}
  \caption{Heat-map matrix of the AUC values in Tab.\,\ref{tab:validation_results}.}
  \label{fig:auc_colorscale}
\end{figure}

Results of the validation process in terms of ROC AUC are reported in Tab.\,\ref{tab:validation_results} and Fig.\,\ref{fig:auc_colorscale}. 

\begin{figure}
  \centering
  \includegraphics[width=.7\textwidth]{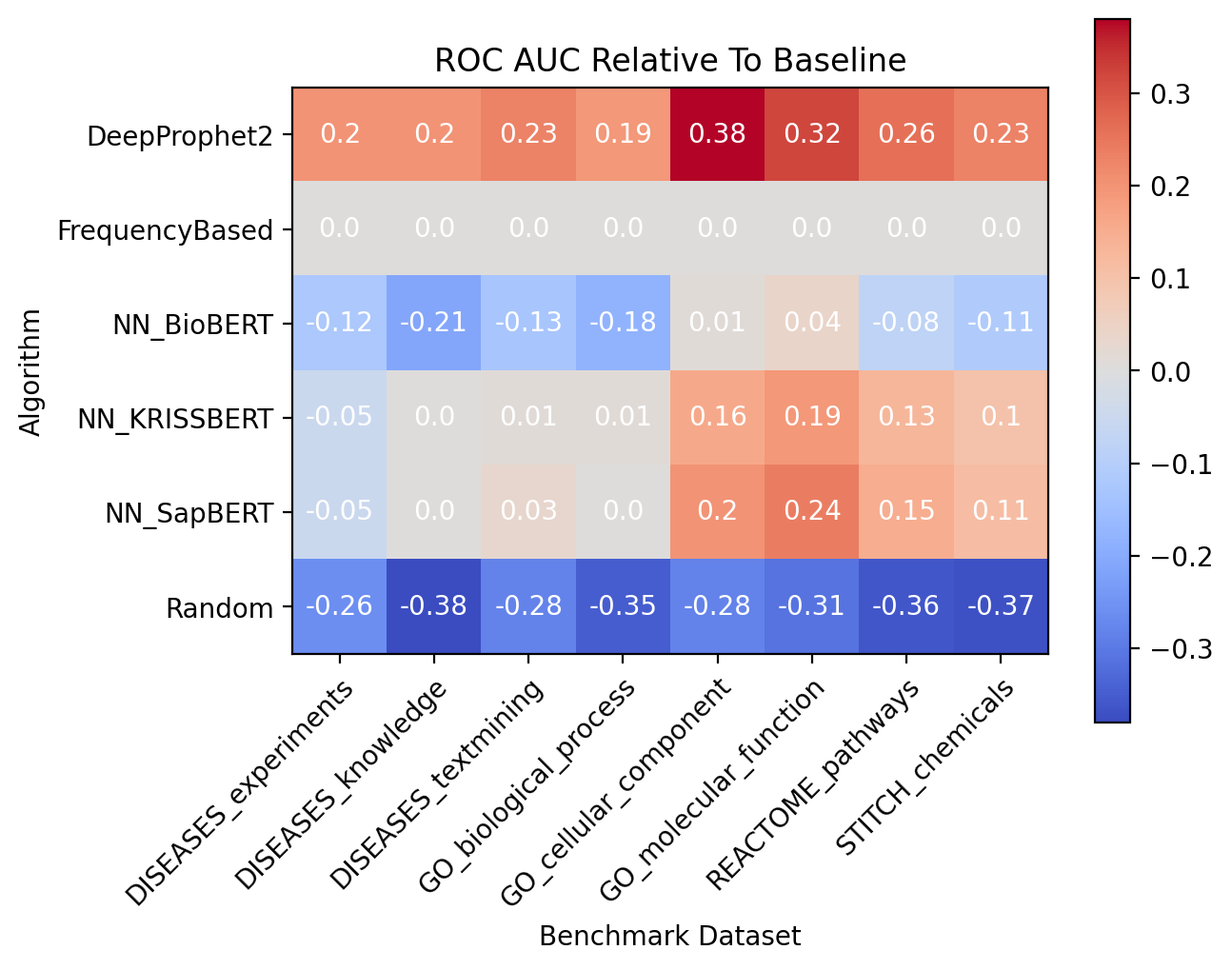}
  \caption{
  Heat-map matrix of the differential ratio of AUC values in Tab.\,\ref{tab:validation_results} w.r.t. the FrequencyBased baseline model. The cell value for each algorithm-dataset pair $(a,d)$ is defined as $r_{ad} = \nicefrac{(\text{AUC}_\text{(a,d)} - \text{AUC}_\text{(f,d)})}{\text{AUC}_\text{(f,d)}} $ where $f$ denotes the FrequencyBased algorithm.}
  \label{fig:auc_colorscale_relative}
\end{figure}

As can be seen, DP2 consistently outperforms all the other algorithms on every benchmark dataset. Looking at the differential ratio with respect to the Frequency Based baseline model (Fig.\,\ref{fig:auc_colorscale_relative}), particularly noteworthy is the fact that the improvement is at least $20\,\%$ for every benchmark dataset, a behaviour that is not shown by any of the other model.

For what concerns the NN algorithms, BioBERT performances are matched, if not surpassed, by the baseline model, showing that the generic pre-training on scientific literature text alone is not enough to reach significant results on the gene recommendation task. The other two algorithms (KRISSBERT and SapBERT) reach better performances, with intermediate AUC values between the baseline and DP2 for some benchmarks and matching the baseline for DISEASES benchmarks and GO biological processes.

\begin{table}
\caption{ROC curve AUC values for every algorithm-dataset combination. Best values for each dataset (columns) are highlighted in bold characters, while the second best is highlighted with italic characters.}
\label{tab:validation_results}
\vspace{3pt}
\centering
\begin{tabular}{@{}l|ccc|ccc|c|c|@{}}
\toprule
\multicolumn{1}{c|}{} &
  \multicolumn{3}{c|}{\textit{\textbf{DISEASES}}} &
  \multicolumn{3}{c|}{\textit{\textbf{GENE ONTOLOGY}}} &
  \textit{\textbf{REACTOME}} &
  \textit{\textbf{STITCH}} \\
\textbf{} &
  \textit{exp} &
  \textit{know} &
  \textit{text} &
  \textit{bio proc} &
  \textit{cell comp} &
  \textit{mol func} &
  \textit{pathways} &
  \textit{chemicals} \\ \midrule
\textit{\textbf{DP2}} &
  \textbf{0.793} &
  \textbf{0.962} &
  \textbf{0.852} &
  \textbf{0.923} &
  \textbf{0.950} &
  \textbf{0.953} &
  \textbf{0.982} &
  \textbf{0.972} \\
\textit{\textbf{NN\_BioBERT}} &
  0.576 &
  0.630 &
  0.597 &
  0.635 &
  0.698 &
  0.751 &
  0.717 &
  0.695 \\
\textit{\textbf{NN\_KRISSBERT}} &
  0.634 &
  \textit{0.803} &
  0.699 &
  \textit{0.784} &
  0.802 &
  0.859 &
  0.879 &
  0.874 \\
\textit{\textbf{NN\_SapBERT}} &
  0.630 &
  0.799 &
  \textit{0.708} &
  0.767 &
  \textit{0.832} &
  \textit{0.892} &
  \textit{0.905} &
  \textit{0.883} \\
\textit{\textbf{MostFreqGenesModel}} &
  \textit{0.662} &
  0.802 &
  0.693 &
  0.768 &
  0.693 &
  0.716 &
  0.780 &
  0.788 \\
\textit{\textbf{RandomModel}} &
  0.491 &
  0.499 &
  0.501 &
  0.500 &
  0.499 &
  0.500 &
  0.498 &
  0.499 \\ \bottomrule
\end{tabular}
\end{table}

\subsection{Optimization and Exploration in the Hyperparameter Space}

\noindent Due to the complexity of the developed model and of the task it tries to solve, it was necessary to perform several optimization procedures in order to obtain a model which could be optimal in the framework of the bias-variance trade-off.
In particular, a first iteration of hyper-parameter tuning was carried out considering as degrees of freedom two parameters that could have the bigger impact on the results:
\begin{enumerate} 
    \item [1.] Embedding size (or network \emph{width}) 
    \item [2.] Number of transformer layers (or network \emph{depth}).
\end{enumerate}
Further analyses will be carried out for future versions of the algorithm.

The role played by these two properties of the network is crucial.
The size of the embedding indeed describes the amount of information that can be preserved during the encoding process \citep{bib:principled_approach}. If too much information is preserved, the encoder is unable to perform a dimensionality reduction that explains the intrinsic properties of biological processes, making the network prone to overfitting. On the other hand, if too much information is destroyed, the model is unable to explore the connection between the statistical units that occur in the training set, leading to underfitting. ~\citep{bib:understanding_dimensionality}.

The depth of the network instead controls the dimension of the hypothesis space in which the possible functions that solve the learning task are found. If the hypothesis space is too large, the network tends to overfit (as it becomes possible to ``interpolate'' the training set), while conversely, a hypothesis space that is too small leads to results that are not enough explanatory of the phenomena under investigation.

\begin{figure}
  \centering
  \setcounter{subfigure}{0}
  \subfloat[][]{\includegraphics[width=0.5\textwidth]{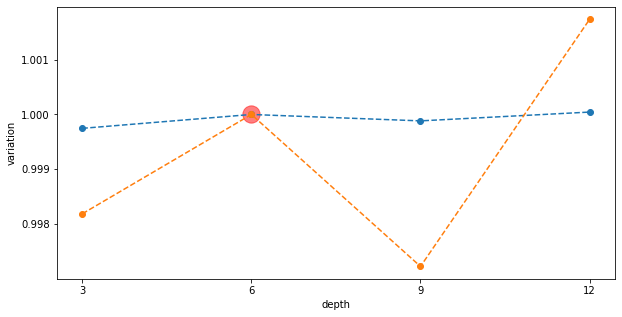}} 
  \subfloat[][]{\includegraphics[width=0.5\textwidth]{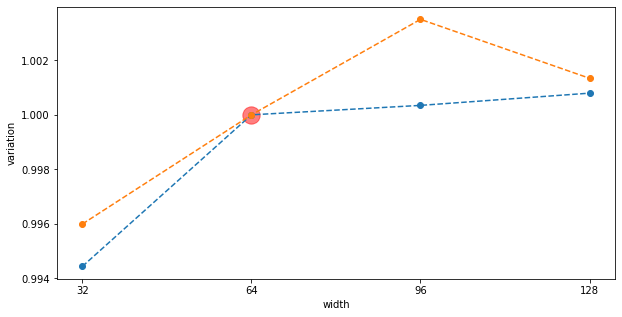}}
  \caption{Ratio between AUC calculated through the validation process described in \ref{sec:validation} with respect to the AUC of the algorithm actually in production, in dependence of different hyperparameters values. This representation is intended to highlight the variation of performance from the point of view of the selected hyper-parameters space point. As can be seen, the chosen point (highlighted with a bigger dot) is a local maximum in the subspace induced by width. In blue: disease completion task, in orange: pathway completion task.}
  \label{fig:auc_depth_width}
\end{figure}

The procedure used to explore the hyperparameter space is outlined as Algorithm \ref{alg:hyperparam_tuning}. In Fig.\,\ref{fig:auc_depth_width} it is shown a representation of the ratio between the AUC, computed during the validation step (more details in paragraph\,\ref{sec:validation}), of two different hyperparameters settings compared with the ones obtained with the algorithm actually in production. As can be seen, sub-optimal solutions were obtained, hence it is planned to define, in the following versions of the project, a more robust optimization procedure (e.g. using a grid search \citep{bib:grid_search}, genetic algorithms or Bayesian optimization \citep{bib:Bayesian_optimization}).

\begin{algorithm}
\caption{Hyperparameter Tuning}
\label{alg:hyperparam_tuning}
\begin{algorithmic}
\State $W \gets \{32,64,96,128\}$  \Comment{Quantization of Width}
\State $D \gets \{3,6,9,12\} $  \Comment{Quantization of Depth}
\State $res[w,d] \leftarrow 0 \hspace{2em} \forall w \in W, d \in D$
\For{$w \in W$}  
    \For{$d \in D$}
        \State $res[w,d] \leftarrow loss_{val}(network(w,d)) $ \Comment{Filling the grid of results}
    \EndFor
\EndFor
\State $w_{best},d_{best} \leftarrow argmin(res) $ \Comment{Selection of the best point according to the procedure}
\end{algorithmic}
\end{algorithm}

\subsection{Coverage}

\noindent In this section, it will be discussed the Coverage \citep{bib:beyond_accuracy}, a metric of high relevance in the evaluation of the model. 

\begin{definition}
(Coverage) Given a recommender system that is trained in order to treat K possible different recommendations, and given $K_{test}$ as the number of different recommendations produced in testing coverage is defined as
\begin{equation}
    C = \frac{K_{test}}{K} .
\label{eq_04} 
\end{equation}
\end{definition}

\noindent This metric provides a very unique perspective on the behaviour of the system from the perspective of a recommender system based on embedding technology. The GeneRecommender system maps the data into a space that preserves the inner structure of the relationships and encodes the information in a particular geometry. The system's query, therefore, works by taking into account neighbourhoods created by the distance from the point of interest (in this case, for example, a gene) in the projection space.

It can be imagined that if neighbourhoods are denser (in terms of contained genes) then more genes are recommended to the final user since the probability of finding a certain gene in a neighbourhood increases with its density \citep{bib:unsupervised_deep_embedding_for_clustering}.

Focusing for a moment on the density of these neighbourhoods: since, according to the notion of semantic distance, the points in the embedding space have distances related to the semantic difference between the genes \citep{bib:distance_metric_learning}, it can be assumed that if the neighbourhoods of points are denser, the system was able to extrapolate a richer meaning from the difference between the genes.

\begin{figure}
  \centering
  \setcounter{subfigure}{0}
  \subfloat[][]{\includegraphics[width=0.5\textwidth]{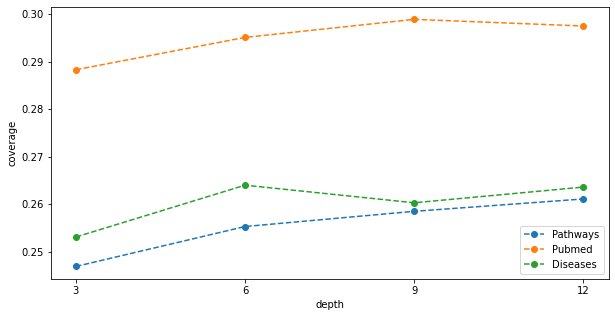}}
  \subfloat[][]{\includegraphics[width=0.5\textwidth]{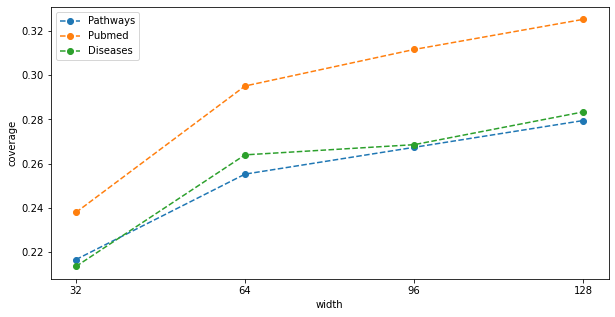}}
  \caption{Dynamics of coverage metric in the space of hyperparameters.
  As it can be seen from the plots, the selected point offers an optimal level of coverage, especially in diseases curve where this point is also a local maxima in the subspace induced by the depth of the network.
  }
  \label{fig:coverage_depth_width}
\end{figure}

Indeed, if points are more clustered while preserving distances that are semantically correct (in the sense that they show consistency with semantic distance), then the system was able to find ways to arrange points that highlight richer links between genes, enhancing the separation between genes with weaker relationships and encouraging the aggregation between the ones that are more correlated. As can be seen from the plot (Fig.\,\ref{fig:coverage_depth_width}), this heuristic seems to be confirmed by the increase of coverage in the function of the depth and width of the network, showing that if the function mapping genes in the embedding space is richer in terms of expressivity an increase in measured coverage is observed.
Indeed, the model deployed on www.generecommender.com has the following parameters: width equal to $64$ and depth equal to $6$. The reason behind these choices is purely based on their economic impact which will not be discussed further. 

\section{Discussion}

\subsection{Use Cases}

\noindent In this section, possible uses of the proposed AI engine will be described, considering some common tasks in various medical and biological research fields that make the recommender system optimal, and also possible situations where this system is less suitable will be highlighted.
For the reader's convenience, the main topics are listed below:
\begin{itemize}
    \item \nameref{par:use_case_insight}
    \item \nameref{par:use_case_rare_diseases}
    \item \nameref{par:complete_existing_pathway}
    \item \nameref{par:use_case_grant}
    \item \nameref{par:use_case_gene_selection}
\end{itemize}

\noindent Before explaining each use case, it is necessary to highlight the fact that, as said in previous paragraphs, the system was created in order to work by emulating a researcher in the life-science field, simulating the reasoning process that brings the researcher from a hypothesis to a discovery. In fact, the following presented use cases should be considered hypothetical, and due to the novelty of the platform other unexplored use cases could be defined in the future. At the current time has not been possible to trace any published paper referring to the platform.

\subsubsection{Insight}
\label{par:use_case_insight}
\noindent 
The algorithm can be seen as an information enhancer: the system takes the knowledge in the genomic field as a starting point and uses it to perform its own discoveries that are then proposed to the final user to suggest new paths to follow during the research.
Deadlocks are extremely common in a high-global-information research topic, and usually, they are not related to the lack of knowledge of the researcher. Indeed, it may happen in fact that a highly qualified researcher may face a deadlock during the research process due to external factors or intrinsic limits of human nature such as a restricted point of view on the problem.  

The system may be beneficial in this situation, that can be defined as an Insight Problem, since it could offer the point of view of the machine, less subject to different kinds of biases that could make human work more difficult.
The algorithm, in fact, tries to find the best point of view to observe every gene (in terms of relationships between other genes), since it has to perform a latent representation of all the genes that encode as much information as possible on the situations that see that gene as a participant.
With a very abstract visualization, it is possible to imagine that while the human point of view tends to be centred in the area of study or research, the system tends to put its point of view in the centre of the gene or the pathway or the disease that it’s being requested. 

\subsubsection{Rare Diseases} 
\label{par:use_case_rare_diseases}

\noindent Secondly, this algorithm can be paramount to getting insights where the amount of available information is limited. Therefore this model could be useful, for example, in the analysis of rare diseases. 
Rare diseases are usually characterized by sparsity of information, in terms of availability and in terms of nature of discoveries, and is therefore difficult for a human researcher to understand the link that connects this kind of disease with one or more genes. This system tries to understand what is the logical connection between genes, giving the researcher an idea of the connections that a disease could have with genes. 
It is important to  notice, anyway, that due to the nature of the model that is proposed, there is a necessity for the researcher to have some (even little) global knowledge on the topic of research.
The mechanism that regulates the system can be assimilated to a multimedia recommender system. If a movie has not been reviewed or seen at least one time the recommender system cannot perform inference on it, since there is a lack of contextualization of the entity that is being proposed.
In the same way, a profound lack of global information in certain research (e.g. extremely rare diseases) is not an optimal context for doing inference with this system. The ``researcher'' still needs sources to study in order to transform information into knowledge.
With some abstraction, the case study relative to rare diseases can be modeled as a situation where there is little global information along with a relatively deep knowledge of the domain in the researcher.

A practical example of the application of GeneRecommender to a case study involving a rare disease is obtained by comparing the results of an RNAseq study \citep{bib:paper_FSHD} regarding a rare disease, Facioscapulohumeral dystrophy (FSHD), with the suggestions of AI. In the study, a cell culture of control myoblasts expressing DUX4 (a transcription factor involved in the pathology) was used to represent the transcriptional changes associated with FSHD. On the other side, GeneRecommender received as input a set of genes with an already known correlation with the pathology (DUX4, FRG1, SMCHD1, FAT1, FSHMD1A, TUBB7P, SMARCA5). It was found that among the first $30$ genes obtained in output, ranked by similarity with input genes, $7$ were also identified as significantly overexpressed in the mentioned study. The algorithm, therefore, could provide meaningful suggestions also in case of rare diseases, at a fraction of the cost of an RNAseq study.

\subsubsection{Complete Existing Pathways}
\label{par:complete_existing_pathway}

\noindent The algorithm could be used also for extending existing pathways with not already known interactions. In this context, the user can insert as input all the genes constituting a pathway to find possible new candidates that extend it. To test the quality of results for this application a simple test was carried out: two different versions of Reactome \citep{bib:reactome} (version 69 of December 2019 and version 76 of March 2021) have been compared to identify the new discoveries introduced in the time between the two. In the meantime, a version of DP2 trained on publications until the 1st of June 2017 has been selected, leaving an interval of two years in which Reactome could undergo its process of revision and update of new knowledge already present at the time of training. The gene sets of the pathways of the oldest version (2019) have been used as input for the algorithm in order to obtain predictions of genes (50 output genes each) that would be added in the future. The aim of this analyses is to quantify the number of matches, defined as a gene introduced in a pathway that has also been predicted by the AI.
In the 2019-2021 interval, 430 pathways were modified in subsequent versions of Reactome, and 202 of them (47\%) have at least a matching with the AI predictions. In the same period, a total of distinct 1846 genes were added to the pathways and 164 of them (9\%) have at least a match in the predictions.

\subsubsection{Grant}
\label{par:use_case_grant}

\noindent The authors of this use case wish to bring attention to a unique scenario in which the GeneRecommender algorithm and platform can be utilized. It is not uncommon for a researcher to identify a potential opportunity for obtaining a public grant in a field that is unfamiliar to them. In such instances, the researcher may be faced with a vast amount of information about the research topic but limited knowledge of the subject.
In traditional circumstances, the researcher would be required to rapidly study a substantial number of papers within a constrained time frame, which may be challenging. However, GeneRecommender and DeepProphet2 are designed to offer support to researchers in such situations. When approaching the definition of a project related to the grant, the system can provide a starting point by suggesting relevant genes, allowing for faster progression and ensuring that all relevant information is readily available.

\subsubsection{Gene Selection Analysis}
\label{par:use_case_gene_selection}
\noindent This ability offers the possibility to reduce the amount of unused information in the analysis of the problem, offering a final perspective on the use cases of the system: Gene Selection Analysis.
The system in fact can optimize automatically, thanks to the notion of similarity among genes, the weight of every gene in terms of information that is related to the case study, allowing the researcher to focus on those that are most relevant to the problem. The most classical use case is to filter the output of a gene expression analysis related to a given disease by using the AI system to highlight the most promising genes. Indeed, over/under expressed genes that are also identified by the system as related to the study might be the ones to focus on. In this way, the algorithm can reduce the amount of targets that are in need for further analysis.

\subsection{Embedding Space Analysis}

\begin{figure}
    \centering
    \setcounter{subfigure}{0}
    \subfloat[][\label{fig:methylation_keratinization_a}]{\includegraphics[width=.5\textwidth]{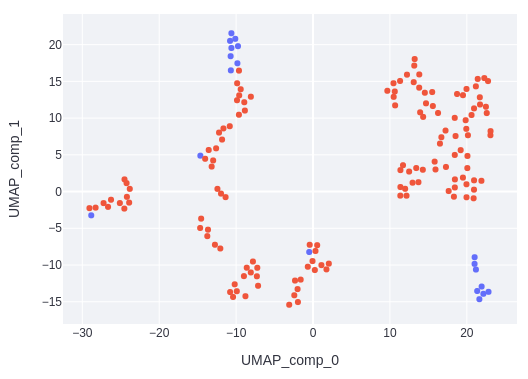}} 
    \subfloat[][\label{fig:keratinization_methylation_b}]{\includegraphics[width=.50\textwidth]{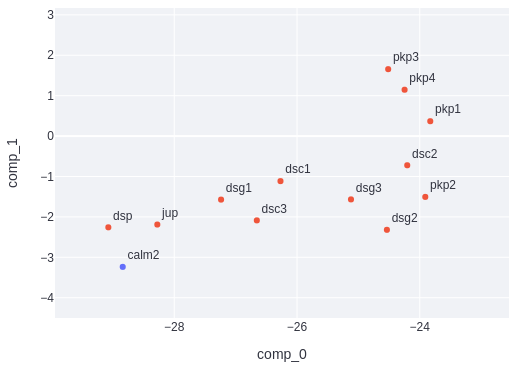}}\\
    \caption{Two-dimensional representation of keratinization and protein methylation pathways (red dots represent genes in the keratinization pathway and blue dots represent genes in the methylation pathway).}
    \label{fig:methylation_keratinization}
\end{figure}

\noindent In Fig.\,\ref{fig:methylation_keratinization} the analysis of the two components projections (obtained with the UMAP algorithm~\citep{bib:UMAP} of gene embedding vectors involved in two different pathways, one for protein methylation and one for keratinization, is displayed.

On the left side of the plot (Fig.\,\ref{fig:methylation_keratinization}\,a) it is possible to see a cluster of genes involved in keratinization, which are all involved in cell adhesion and in particular are part of desmosomes (DSG family, PKP family, DSC family, DSP, JUP)~\citep{bib:desmosomes}. 

This cluster also contains a gene associated with protein methylation, CALM2. This gene belongs to the calmodulin family~\citep{bib:calmodulin}, a group of proteins that bind $\text{Ca}2^+$ and play a role in signalling pathways, cell cycle progression, and proliferation. Many of the proteins encoded by the aforementioned genes, such as DSG1, DSG3, DSC1, and DSC3, bind $\text{Ca}2^+$ as well. Some cluster proteins, such as the one encoded by PKP2, do not directly bind $\text{Ca}2^+$, but rather maintain transcription of genes that control intracellular calcium cycling~\citep{bib:pkp2}. A representation of Gene set enrichment made with GO is present in Appendix \ref{app:integrative_plots}, Fig.\,\ref{fig:embedding_ca_enrichment} and in details in Appendix \ref{app:integrative_tables}, Tab.\,\ref{tab}.

This empirical test shows that the network has learned to encode information about cell adhesion and $\text{Ca}2^+$ dependent signalling pathways in the embedding of genes. These features, plus others unexplored, have been sufficient to represent a well characterised cluster (qualitatively defined) also after the reduction in two dimensions. In the original embedding space, with many more dimensions, the information encoded could be, in principle, much more.

\begin{figure}[p]
  \centering
  \includegraphics[width=0.95\textwidth]{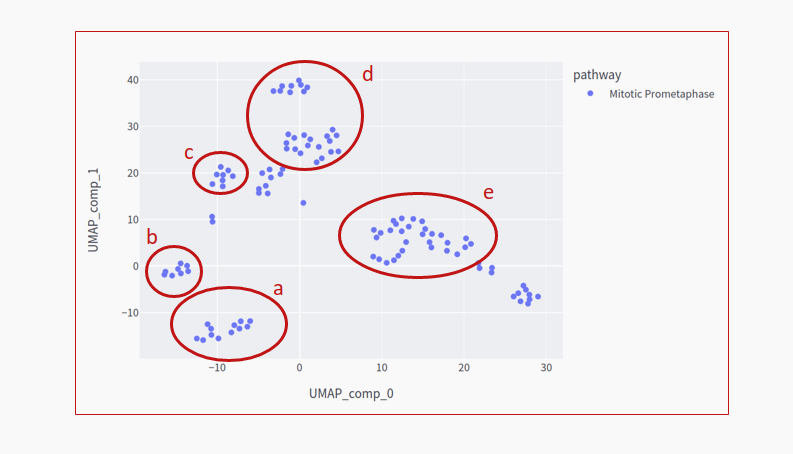}
  \caption{2-D representation of the embeddings of genes constituting a single pathway.}
  \label{mitotic_prometaphase_pathway}
\end{figure}

\begin{figure}[p]
  \centering
  \includegraphics[width=0.95\textwidth]{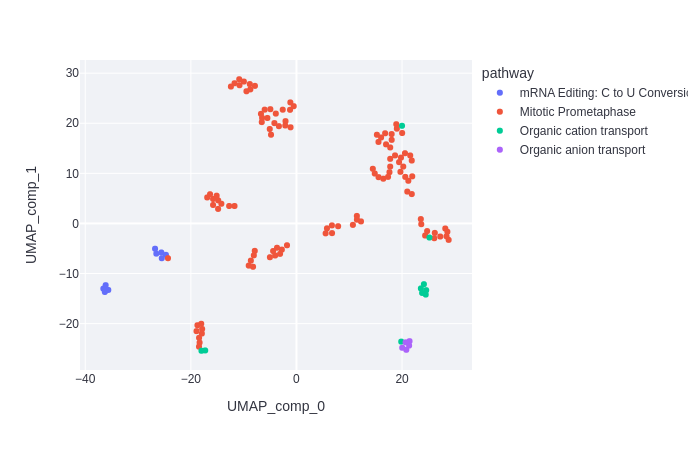}
  \caption{Representation of the 2-D embeddings of 4 different pathways.}
  \label{4_distinct_pathways}
\end{figure}

Other pathways were investigated with different aims. For example, in Fig.\,\ref{mitotic_prometaphase_pathway} the reader can find a projection plot of a single pathway, with the purpose of checking qualitatively if the clusters ideally generated in the projected space contain some biological information. Some of them:  a: Group of tubulin $\alpha$ and $\beta$, they form the heterodimer constituting microtubules. b: This genes (PPP2R1A, PPP2R5D, PPP2R5A etc.) encode for subunits of the phosphatase 2A implicated in the negative control of cell division. c: This area contains genes involved in the formation of nuclear pore complex~\citep{bib:NPC} like NUP family and SEC13, SEH1L. d: This area contains genes involved in the centromeric complex formation~\citep{bib:centromeric_complex}, essentials for proper kinetochore function (CENPP, CENPN, CENPL, ZWILCH, MIS12, SPC25, NSL1, SGO2 etc.). e: In this cluster are present genes involved in the formation of spindle  assembly checkpoint~\citep{bib:spindle_assembly_checkpoint} (a control that prevents an uncorrected separation of chromosomes) and chromosome segregation (NUF2, KNL1, DYNCH1H1, KIF2C, BUB1, KIF2A, etc.) 
Fig.\,\ref{4_distinct_pathways} instead investigates the separation between different pathways in the projected space. Could be seen a clear separation of four distinct pathways, the two of them with a similar biological function, organic anion transport and organic cation transport are on the same vertical line, as if the information embedded in the $\text{comp}_0$ conveyed part of that information. Genes involved in mRNA editing otherwise are in a well separated area of the plot.

Analysis of this kind can be considered as shreds of evidence that the DeepProphet2 transformer model is able to extract biological relationships between genes. By utilizing this approach, pathways previously unknown to the net are understood fully (without being included in the training process) by the system. As a result of these experiments, it is possible to conclude with sufficient confidence that the developed model can predict new candidates to be included in a particular pathway.

\subsection{Conclusion}

%parte 1. Conclusioni sull'avanzamento tecnologico
\noindent From a technological point of view, this system could offer a different perspective on the task of genetic (or more in general biomedical) recommendation.
Transformer technology was not simply used to produce results, but was also adapted in a way to be an optimal instrument to deal with the task of interest.
It is important to address the fact that the system is not meant to be a replacement for human researchers' activity: the objective of the work is to produce a system that works in synergy with the researcher, emulating their way of working to produce results that can be justified and understood in the most proper way.
%parte 2. Possibili risvolti nel campo dell'informatica medica
It is hoped that this approach to research problems may serve as a preliminary step towards the development of new methods for making discoveries. Rather than viewing the machine as merely a tool for storing and retrieving data, it is utilized as a means for uncovering patterns that may be difficult to detect through human observation alone.
%parte 3. Considerazioni su possibili
Like a human researcher, the system is able to analyse and learn to improve itself and its reasoning capabilities. Aside from this ability, future research will be performed in order to improve the algorithms that sustain the system, following the global improvement of machine learning techniques in order to always have state-of-the-art prediction quality levels.

\section*{Acknowledgments}
We thank Dr. Luigi Bondurri, Dr. Alessandro Radaelli, and Dr. Christian Lautenschläger for their valuable discussions and support.

\noindent The ``Tech Fast'' grant from Regione Lombardia made all this work possible.

\clearpage
\appendix

\section{Integrative Tables}
\label{app:integrative_tables}

\begin{table}[h]
\centering
\caption{\label{demo-table}Gene set molecular function enrichment (GO) of genes presented in Fig.\,\ref{fig:methylation_keratinization}\,b composing protein methylation and keratinization pathways.}
\vspace{3pt}
\label{tab}
\renewcommand{\arraystretch}{2}
\begin{tabular}{p{0.75cm}>{\raggedright}p{0.3\textwidth}p{1.8cm}p{1.8cm}p{1.8cm}p{1.8cm}}
\toprule
\textbf{Index} & \textbf{Name} & \textbf{P-value} & \textbf{Adjusted p-value} & \textbf{Odds Ratio} & \textbf{Combined score} \\ 
\midrule
1 & cell adhesive protein binding involved in bundle of His cell-Purkinje myocyte communication (GO:0086083) & 4.83E-17 & 1.35E-15 & 99935 & 3754562.93 \\
2 & protein binding involved in heterotypic cell-cell adhesion (GO:0086080) & 1.21E-14 & 1.70E-13 & 2497.75 & 80033.17 \\
3 & calcium ion binding (GO:0005509) & 7.13E-10 & 6.66E-09 & 67.22 & 1415.64 \\
4 & adenylate cyclase binding (GO:0008179) & 6.48E-03 & 2.12E-02 & 184.98 & 932.06 \\
5 & calcium channel inhibitor activity (GO:0019855) & 7.13E-03 & 2.12E-02 & 166.47 & 822.99 \\
6 & metal ion binding (GO:0046872) & 1.11E-08 & 7.78E-08 & 44.56 & 816.06 \\
7 & protein kinase C binding (GO:0005080) & 3.80E-04 & 1.77E-03 & 84.33 & 664.11 \\
8 & protein phosphatase activator activity (GO:0072542) & 8.42E-03 & 2.12E-02 & 138.72 & 662.67 \\
9 & titin binding (GO:0031432) & 8.42E-03 & 2.12E-02 & 138.72 & 662.67 \\
10 & phosphatase activator activity (GO:0019211) & 9.06E-03 & 2.12E-02 & 128.04 & 602.21 \\
\bottomrule
\end{tabular}
\end{table}

\vspace{3cm}

\pagebreak
\clearpage

\section{Integrative Plots}
\label{app:integrative_plots}
\begin{figure}[h]
    \centering
    \setcounter{subfigure}{0}
    \subfloat[][]{\includegraphics[width=.40\textwidth]{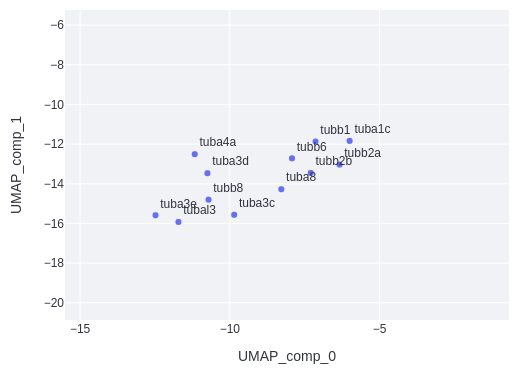}} \quad
    \subfloat[][]{\includegraphics[width=.40\textwidth]{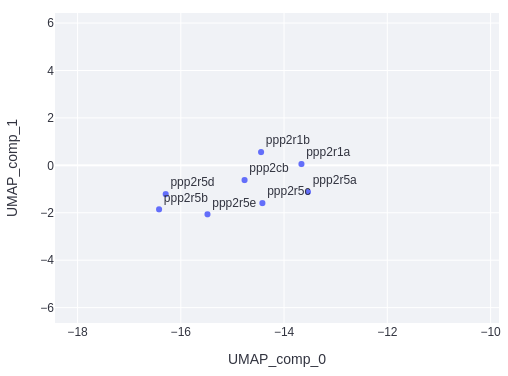}} \\
    \subfloat[][]{\includegraphics[width=.40\textwidth]{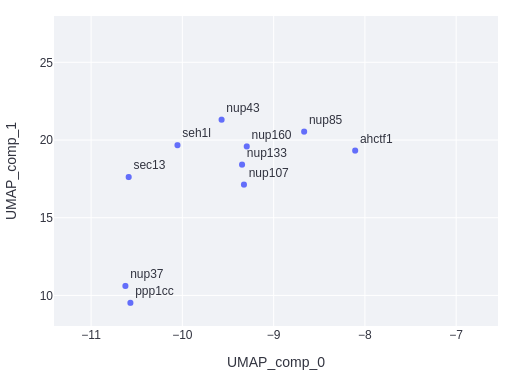}} \quad
    \subfloat[][]{\includegraphics[width=.40\textwidth]{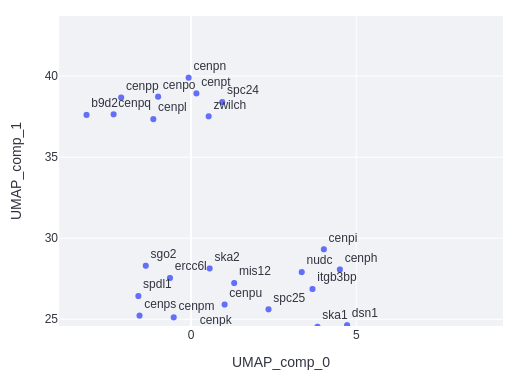}} \\
    \subfloat[][]{\includegraphics[width=.40\textwidth]{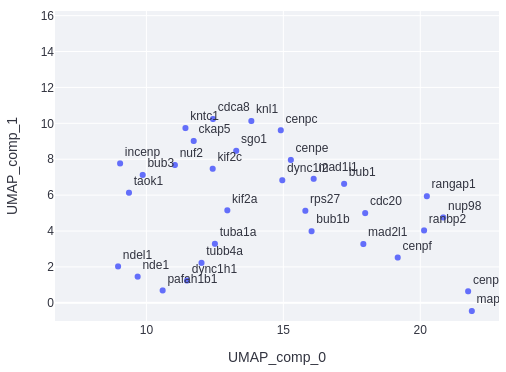}}
    \caption{Zoom-in of the clusters identified in Fig.\,\ref{mitotic_prometaphase_pathway}. }
    \label{fig: Integrative plot}
\end{figure}

\begin{figure}[h]
  \centering
  \includegraphics[width=0.8\textwidth]{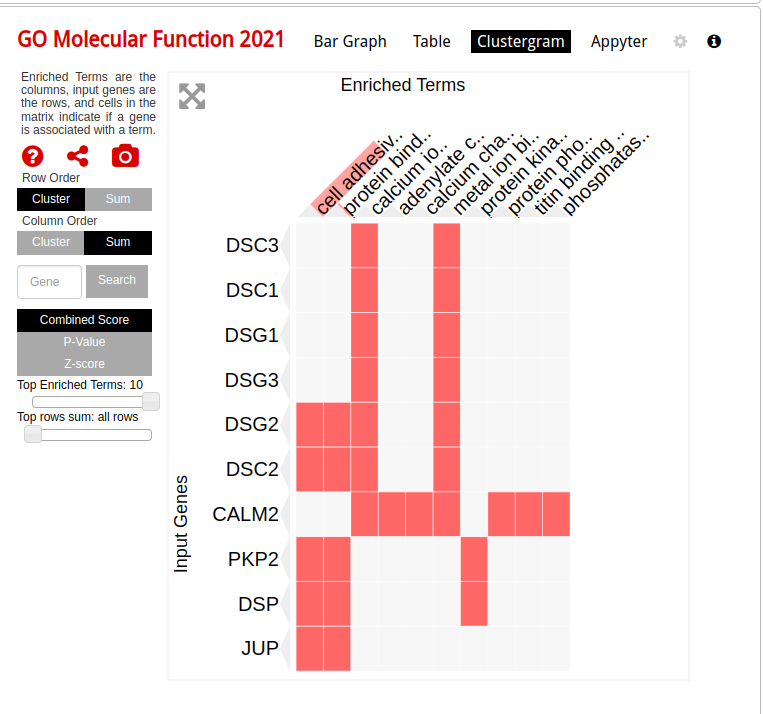}
  \caption{Clustergram of gene set enrichment of GO molecular function, source: Enrichr.}
  \label{fig:embedding_ca_enrichment}
\end{figure}

\pagebreak
\clearpage
%Bibliography
\bibliographystyle{abbrvnat}
\bibliography{references}  

\end{document}